\documentclass[reprint, aps,pra,twocolumn,superscriptaddress]{revtex4-1}
\usepackage{braket}
\usepackage{amssymb}
\usepackage{amsmath}
\usepackage{epsfig}
\usepackage{color}
\usepackage{graphics, graphicx}
\usepackage{bbold}
\usepackage{psfrag}
\usepackage{mathcomp}
\usepackage{subfigure}
\usepackage{verbatim}
\usepackage[colorlinks, citecolor=blue]{hyperref}
\usepackage[normalem]{ulem}
\usepackage{bm}
\def\cp#1{\mathbf{#1}}

\begin{document}

\title{Rabi-coupling-induced three-component quantum droplet in ultracold Bose gases}
\author{Xiao Ding}
\affiliation{Beijing National Laboratory for Condensed Matter Physics, Institute of Physics, Chinese Academy of Sciences, Beijing 100190, China}
\affiliation{School of Physical Sciences, University of Chinese Academy of Sciences, Beijing 100049, China}
\author{Dajun Wang}
\affiliation{Department of Physics, The Chinese University of Hong Kong, Hong Kong, China}
\author{Xiaoling Cui}
\affiliation{Beijing National Laboratory for Condensed Matter Physics, Institute of Physics, Chinese Academy of Sciences, Beijing 100190, China}

\date{\today}

\begin{abstract}
We uncover a mechanism for realizing three-component quantum droplets in ultracold Bose gases, where only one interspecies interaction is attractive. In this scheme, the interspecies attraction leads to a self-bound binary droplet, and the third component is incorporated through Rabi coupling with one component of the binary droplet. We find that a stronger Rabi coupling leads to a larger fraction of the third component, but also destabilizes the entire droplet due to the involvement of more repulsive forces. Such instability can be remedied by a finite detuning between the Rabi-coupled components. We demonstrate these results in realistic Na-Rb mixtures, using both thermodynamic analyses and numerical simulations based on extended Gross-Pitaevskii equations. Our work outlines a general route for stabilizing multicomponent droplets by bridging an existing binary droplet with additional components via suitable single-particle fields.
\end{abstract}
\maketitle

\section{Introduction}

Quantum droplets in ultracold atomic gases represent a novel state of matter characterized by their self-bound nature, where beyond-mean-field quantum fluctuations stabilize the system against collapse~\cite{Petrov_1}. In recent years, quantum droplets have been successfully realized in both dipolar Bose gases~\cite{Pfau_1, Pfau_2, Ferlaino_1, Modugno_1, Pfau_3, Ferlaino_2} and alkali-metal Bose-Bose mixtures~\cite{Tarruell_1, Tarruell_2, Fattori_1, Fort_1, Fort_2, Wang_1, Modugno_2, Burchianti}. In particular, a self-bound droplet of binary bosons has been shown to exhibit fascinating properties distinct from a repulsive Bose gas, including collective excitations~\cite{Petrov_1, Petrov_2, Liu, Reimann_1, Citro, Zhang_1}, low-dimensional behaviors~\cite{Astrakharchik, Giorgini_1, Zhang_2, Ma, Mistakidis_1}, liquid-gas transition and coexistence~\cite{Giorgini_2, Cui_1, Yu}, vortex structure~\cite{Malomed_1, Li, Reimann_2, Ancilotto, Cui_2, Oktel_1}, nonequilibrium dynamics~\cite{Malomed_2, Fattori_2, Boronat, Modugno_3, Cui_3, Schmelcher}, and interplay with spin-dependent fields~\cite{Malomed_3, Bourdel, Cui_4, Oktel_2, Cui_5, Mazzanti, Yin}, etc.

Beyond the framework of binary droplets, an interesting cutting-edge direction is droplet formation with multiple components. As a typical example, three-component droplets have been explored in both three dimensions (3D)~\cite{Cui_6, Cui_7, Macri} and one dimension (1D)~\cite{Mistakidis_2, Mistakidis_3}, where intriguing quantum phases, such as Borromean droplets~\cite{Cui_6} and shell-shaped droplets~\cite{Cui_7}, have been revealed due to the interplay of mean-field interactions, quantum fluctuations, and spin degrees of freedom. Nevertheless, for three components to bind together as a droplet in 3D free space, at least two interspecies couplings must be attractive~\cite{Cui_6, Cui_7, Macri}. In practice, this requirement is very stringent, since in most realistic ultracold boson mixtures only one interspecies coupling can be tuned to be attractive at a given time. A natural question then arises: is it possible to bind three-component bosons into a droplet when attraction exists in only one interspecies channel? The present work aims to address this question.

\begin{figure} [h]
\includegraphics [width=6.5cm]{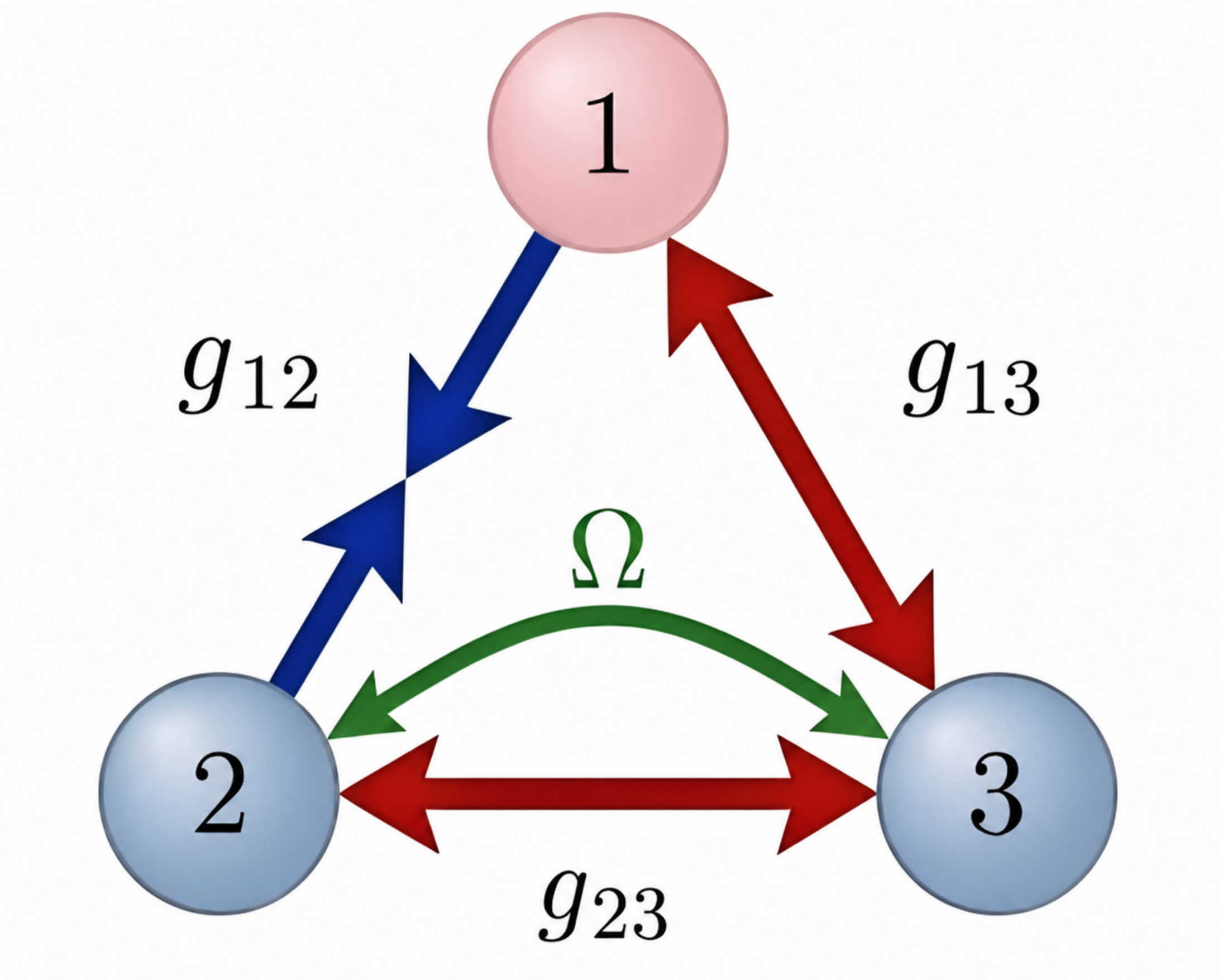}
\caption{Illustration for achieving a three-component (1,2,3) quantum droplet. Components 1 and 2 form a binary droplet in vacuum due to interspecies attraction, and component 3 joins via coherent Rabi coupling with component 2, thereby facilitating the formation of a (1,2,3) droplet. The coupling constants in all other interaction channels are repulsive.}
\label{fig_schematic}
\end{figure}

Here, we propose a mechanism for realizing three-component quantum droplets in ultracold boson mixtures with only one attractive interspecies coupling. Our scheme is illustrated in Fig.~\ref{fig_schematic}: for three-component (1,2,3) bosons, (1,2) form a binary droplet due to interspecies attraction, and the third component (3) joins via coherent Rabi coupling with one component (2) of the binary droplet. In this way, all three components can bind together as a self-bound droplet. Taking a realistic Na-Rb mixture as an example, we verify this proposal using both thermodynamic analyses and numerical simulations based on extended Gross-Pitaevskii equations. Interestingly, we find that a stronger Rabi coupling leads to a larger fraction of component 3, which, on the other hand, destabilizes the entire droplet due to the involvement of more repulsive forces. Such instability can be rescued by a finite detuning between the Rabi-coupled components (2 and 3). These results can be readily tested in current cold-atom experiments. Beyond the three-component droplet, our work also outlines a general route for stabilizing multicomponent droplets---namely, by bridging an existing binary droplet with additional components via suitable single-particle fields.

The remainder of this paper is organized as follows. Section~\ref{model} presents our basic theoretical model. Section~\ref{mf} is devoted to mean-field analysis. Section~\ref{droplet} presents the main results on three-component quantum droplets, including their formation and stability in the thermodynamic limit, verification via extended GP equations, and the effect of magnetic detuning. Section~\ref{summary} provides the summary and outlook of this work.

\section{ Model } \label{model}

We write down the Hamiltonian of three-component (1,2,3) bosons in three dimensions as illustrated in Fig.~\ref{fig_schematic} ($\hbar=1$):
\begin{eqnarray}
H&=&\int d{\cp r} \left[ H_0({\cp r})+U({\cp r}) \right], \label{H} \\
H_0&=& -\sum_{i=1}^3\psi_{i}^\dagger \frac{\nabla^2}{2m_i} \psi_{i} 
-\Omega(\psi_{2}^\dagger \psi_{3}+\mathrm{H.c.}) \\ \delta (\psi_{2}^\dagger \psi_{2}-\psi_{3}^\dagger \psi_{3}) ;\nonumber\\
U&=&\frac{1}{2} \sum_{i,j=1}^3g_{ij}\psi_{i}^\dagger\psi_{j}^\dagger \psi_{j}\psi_{i}. \nonumber
\end{eqnarray}
Here we have simplified the local field operators $\{\psi_{i}^\dagger(\mathbf{r}), \psi_{i}(\mathbf{r})\}$ as $\{\psi_{i}^\dagger, \psi_{i}\}$, which respectively create and annihilate a component-$i$ boson with mass $m_i$ at coordinate ${\cp r}$; $\Omega$ and $\delta$ are respectively the strengths of Rabi coupling and detuning between components 2 and 3; and $g_{ij}=2\pi a_{ij}/m_{ij}$ is the interaction strength between components $i$ and $j$, with scattering length $a_{ij}$ and reduced mass $m_{ij}\equiv m_im_j/(m_i+m_j)$. In this work, we consider all couplings to be repulsive except $g_{12}$, which is sufficiently negative to support a binary (1,2) droplet in vacuum~\cite{Petrov_1}.

As a specific example, we consider a realistic ultracold system of a $^{23}$Na-$^{87}$Rb mixture. Namely, we take components 1 and 2 as the hyperfine states $|F=1,m_{F}=1\rangle$ of $^{23}$Na and $^{87}$Rb atoms, respectively, which features an interspecies Feshbach resonance at $B = 347.65$ G. These two components can support a binary droplet, which has been successfully observed in experiment~\cite{Wang_1}. For component 3, we take it as another hyperfine state $|F=1,m_{F}=0\rangle$ of $^{87}$Rb atom, which can be coupled to component 2 via a Rabi field. Throughout this paper, we consider the $^{23}$Na-$^{87}$Rb mixture near $B\sim 349.28$ G, with fixed scattering lengths $(a_{11}, a_{22}, a_{33}, a_{12}, a_{13}, a_{23})=(60.1,100.1,100.9,-123.2,89.7,100.4)a_0$ ($a_0$ is the Bohr radius)~\cite{Wang_1, Wang_2}. In this regime, we have $g_{12}<-\sqrt{g_{11}g_{22}}$ and thus the (1,2) droplet can be supported in vacuum. Moreover, we consider flexible strengths of $\Omega$ and $\delta$, which can be tuned conveniently by the intensity and frequency of the Rabi field.

\section{Mean-field analysis} \label{mf} 

In this section, we analyze the mean-field stability of three-component bosons against density fluctuations. 
First, under the mean-field treatment, we replace the field operators by \(\psi_{2} = \sqrt{n_{23}}\sin{\frac{\phi}{2}}\), \(\psi_{3} = \sqrt{n_{23}}\cos{\frac{\phi}{2}}\), and \(\psi_{1} = \sqrt{n_{1}}\). Here we have assumed homogeneous densities $n_i\ (i=1,2,3)$, and $n_{23}\equiv n_2+n_3$. The density imbalance between components \(2\) and \(3\) is characterized by the polarization
\(S \equiv (n_{3}-n_{2})/n_{23} = \cos\phi\).  Then we have the mean-field energy density
\begin{equation}
\begin{aligned}
    \mathcal{E}_{\rm{MF}} = & - (\Omega \sqrt{1-S^{2}} + \delta S )n_{23} + \Bigg[ \frac{g_{22}+g_{33}-2g_{23}}{8}S^{2} \\
                            & - \frac{g_{22}-g_{33}}{4}S 
                       + \frac{g_{22}+g_{33}+2g_{23}}{8} \Bigg] n_{23}^{2} \\
                    & +\frac{1}{2}g_{11}n_{1}^{2} + \Bigg[ \frac{g_{13}-g_{12}}{2}S + \frac{g_{13}+g_{12}}{2} \Bigg] n_{1}n_{23}. 
\end{aligned}
\label{E_mf_0}
\end{equation}
The equilibrium polarization \(S\), as a function of \(n_{1}\) and \(n_{23}\), is determined by
\begin{equation}
    \frac{\partial \mathcal{E}_{\rm{MF}}}{\partial S} = 0,
\end{equation}
which further gives $\mathcal{E}_{\rm{MF}}$ as a function of \(n_{1}\) and \(n_{23}\). 

The mean-field stability can be examined by evaluating the second-order variation of $\mathcal{E}_{\rm{MF}}$ with respect to small density fluctuations:
\begin{equation}
\delta^{2}\mathcal{E}_{\rm MF} = \frac{1}{2}\sum_{\alpha\beta} g_{\alpha\beta}^{\rm eff}\, \delta n_{\alpha}\,\delta n_{\beta}, \label{E_mf}
\end{equation}
where \(\alpha,\beta = \{1,23\}\) and the effective interaction strength is defined as 
\begin{equation}
g_{\alpha\beta}^{\rm{eff}} = \frac{\partial^{2} \mathcal{E}_{\rm MF}}{\partial n_{\alpha}\,\partial n_{\beta}}.
\end{equation}
Diagonalization of \(\delta^{2}\mathcal{E}_{\rm MF}\) in Eq.~(\ref{E_mf}) yields the eigenfluctuation modes, from which one can determine the mean-field stability of the system. Notably, the presence of Rabi coupling $(\Omega\neq 0)$ gives rise to a highly nonlinear dependence of $\mathcal{E}_{\rm{MF}}$ on $\{n_{1}, n_{23}\}$, thereby leading to density-dependent stability even at the mean-field level.

Explicitly, Eq.~(\ref{E_mf}) can be diagonalized as
\begin{equation}
\delta^{2}\mathcal{E}_{\rm MF} = \tilde{g}_{1}\, \delta \tilde{n}_{1}^{2} + \tilde{g}_{23}\, \delta \tilde{n}_{23}^{2}, \label{E_mf_2}
\end{equation}
where $\tilde{n}_{1}$ and $\tilde{n}_{23}$ are the two eigenmodes of fluctuations, and $\tilde{g}_{1}$ and $\tilde{g}_{23}$ are their corresponding eigenvalues. A stable mean-field phase corresponds to all these modes having positive eigenvalues, i.e., $\tilde{g}_{1}>0, \tilde{g}_{23}>0$ in Eq.~(\ref{E_mf_2}) and $g_{\alpha\alpha}^{\rm eff}>0$ in Eq.~(\ref{E_mf}). If any of these conditions breaks down, the system tends to become unstable against density fluctuations.

\begin{figure} [t]
\includegraphics [width=\linewidth]{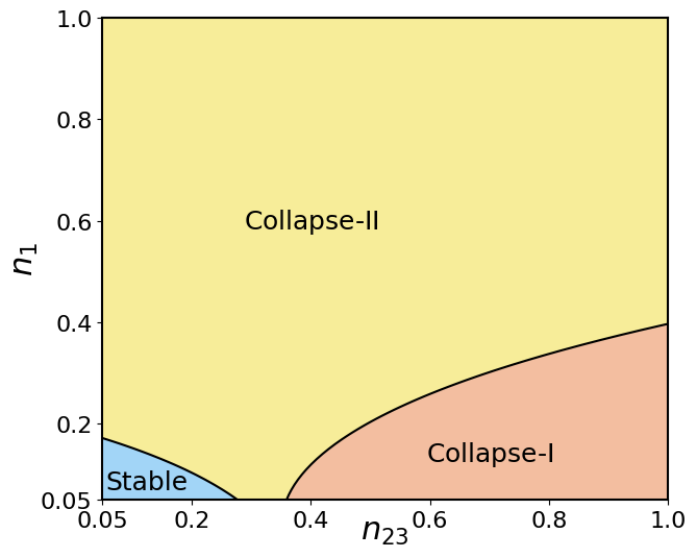}
\caption{Mean-field phase diagram of three-component bosons in the \((n_{1}, n_{23})\) plane. Three phases are identified: a stable miscible phase (``Stable''), and two unstable phases towards collapse of component 1 (``Collapse-I'') and collapse of all components (``Collapse-II''). Here we take the Na-Rb mixture at $\delta=0$, and the densities are scaled by \(\Omega/g_{33}\).}
\label{fig_diagram}
\end{figure}

In Fig.~\ref{fig_diagram}, we map out the mean-field phase diagram in the $(n_{1}, n_{23})$ plane at typical values of $\Omega$ and $\delta=0$. Apart from the mean-field stable regime (marked ``stable''), we find two unstable regions towards mean-field collapse. Specifically, ``Collapse-I'' corresponds to $g_{11}^{\rm eff}<0$, which gives the collapse of component 1, and ``Collapse-II'' corresponds to $\tilde{g}_{1}\tilde{g}_{23}<0$ while $g_{\alpha\alpha}^{\rm eff}>0$, which gives the collapse of all three components.

Starting from the stable region at small $n_{1}$ and $n_{23}$, we can see that increasing $n_{23}$ will drive the system to `Collapse-I', while increasing $n_{1}$ or both densities drives it to `Collapse-II'. This can be understood as follows. In the low-density limit of $(n_{1},n_{23})$, the single-particle physics dominates and the whole system is stable because all $g_{ii}>0$. 
For a fixed small $n_1$, increasing $n_{23}$ introduces a correction to the effective 1-1 coupling strength: $\Delta g_{11}=\frac{g_{13}-g_{12}}{2}\frac{\partial S}{\partial n_{1}} n_{23}$, as seen from the mean-field energy in Eq.~(\ref{E_mf}). Because the 1-2 attraction favors component 2 over component 3, we have $\frac{\partial S}{\partial n_{1}}<0$, making this correction negative and finally leading to $g_{11}^{\rm eff}=g_{11}+\Delta g_{11}<0$, i.e., the Collapse-I region. As both $n_{1}$ and $n_{23}$ continue to increase, the interaction energy becomes dominant, and due to the strong 1-2 attraction the system is driven into the Collapse-II region, where all three components undergo mean-field collapse. This implies the formation of a three-component droplet when the Lee-Huang-Yang (LHY) energy from quantum fluctuations is included, as discussed in the next section.

Note that the mean-field phase diagram in Fig.~\ref{fig_diagram} requires both $n_{1}$ and $n_{23}$ to be finite and therefore does not apply to the cases $n_{1}=0$ or $n_{23}=0$. This is because, in the absence of component 1 [or $(2,3)$], the system is solely composed of $(2,3)$ (or 1) atoms, and the interaction-induced correlation between different components vanishes completely, thereby invalidating the above analysis.

\section{Three-component  quantum droplet}  \label{droplet} 

We now proceed with the formation of three-component droplets by incorporating the effect of quantum fluctuations. Based on the standard Bogoliubov theory, we expand the field operator as
\begin{equation}
    \psi_{i} ({\cp r}) = \sqrt{n_{i}} + \frac{1}{\sqrt{V}} \sum_{k \neq 0} e^{i\textbf{kr}} a_{i\textbf{k}},
\end{equation}
where \(a_{i\textbf{k}}\) is the fluctuation operator for component-\(i\) bosons at momentum \({\cp k}\). The Hamiltonian in Eq.~(\ref{H}) can then be reduced to a bilinear form:
\begin{widetext}
\begin{equation}
\begin{aligned}
    \frac{H}{V} = \mathcal{E}_{\rm{MF}} + & \frac{1}{V} \sum_{{\cp p}\neq {\cp 0}}' \Bigg[ \phi^{\dagger} h_{\textbf{p}} \phi
    - \sum_{i=1}^3 (\epsilon_{i\textbf{p}} + n_{i} g_{ii}) - \Omega \frac{n_{2} + n_{3}}{\sqrt{n_{2}n_{3}}} \Bigg]
\end{aligned}
\end{equation}
where \(\sum'_{{\cp p}}\) denotes the summation only in half of momentum space; \(\mathcal{E}_{\rm{MF}} = \frac{1}{2}\sum_{i,j}g_{ij}n_{i}n_{j} - 2\Omega \sqrt{n_{2}n_{3}} + \delta(n_{2} - n_{3})\), \(\epsilon_{i\textbf{p}} = {\cp p}^{2}/2m_{i}\), \(\phi = \left(a_{1\textbf{p}} \ a_{2\textbf{p}} \ a_{3\textbf{p}} \ a_{1-\textbf{p}}^{\dagger} \ a_{2-\textbf{p}}^{\dagger} \ a_{3-\textbf{p}}^{\dagger}\right)^{T}\) and
\begin{equation}
\begin{aligned}
    h_{\textbf{p}} =
    \scalebox{0.95}{$
    \begin{pmatrix} 
        \epsilon_{1\textbf{p}} + n_{1} g_{11} & g_{12}\sqrt{n_{1}n_{2}} & g_{13}\sqrt{n_{1}n_{3}} & n_{1}g_{11} & g_{12}\sqrt{n_{1}n_{2}} & 
        g_{13}\sqrt{n_{1}n_{3}} \\
        g_{12}\sqrt{n_{1}n_{2}} & \epsilon_{2\textbf{p}} + \Omega \sqrt{\frac{n_{3}}{n_{2}}} + n_{2} g_{22} &  g_{23}\sqrt{n_{2}n_{3}} - \Omega & g_{12}\sqrt{n_{1}n_{2}} & n_{2}g_{22} &
        g_{23}\sqrt{n_{2}n_{3}} \\
        g_{13}\sqrt{n_{1}n_{3}} & g_{23}\sqrt{n_{2}n_{3}} - \Omega &
        \epsilon_{3\textbf{p}} + \Omega \sqrt{\frac{n_{2}}{n_{3}}} + n_{3} g_{33} & g_{13}\sqrt{n_{1}n_{3}} &
        g_{23}\sqrt{n_{2}n_{3}} & n_{3}g_{33} \\ 
        n_{1}g_{11} & g_{12}\sqrt{n_{1}n_{2}} & 
        g_{13}\sqrt{n_{1}n_{3}} & \epsilon_{1\textbf{p}} + n_{1} g_{11} & g_{12}\sqrt{n_{1}n_{2}} & g_{13}\sqrt{n_{1}n_{3}} \\
        g_{12}\sqrt{n_{1}n_{2}} & n_{2}g_{22} &
        g_{23}\sqrt{n_{2}n_{3}} & g_{12}\sqrt{n_{1}n_{2}} & 
        \epsilon_{2\textbf{p}} + \Omega \sqrt{\frac{n_{3}}{n_{2}}} + n_{2} g_{22} &  g_{23}\sqrt{n_{2}n_{3}} - \Omega \\
        g_{13}\sqrt{n_{1}n_{3}} & g_{23}\sqrt{n_{2}n_{3}} & n_{3}g_{33} &
        g_{13}\sqrt{n_{1}n_{3}} & g_{23}\sqrt{n_{2}n_{3}} - \Omega &
        \epsilon_{3\textbf{p}} + \Omega \sqrt{\frac{n_{2}}{n_{3}}} + n_{3} g_{33} \\ 
    \end{pmatrix}.
    $}
\end{aligned}
\end{equation}
The Bogoliubov modes \(\{E_{i\mathbf{p}}\}\) can be obtained by diagonalizing matrix \(\Gamma h_{\mathbf{p}}\), with $\Gamma = \mathrm{diag}(1, 1, 1, -1, -1, -1)$.
Finally, we arrive at the LHY energy density
\begin{equation}
\begin{aligned}
    \mathcal{E}_{\mathrm{LHY}} 
    & = \frac{1}{2V} 
   \sum_{{\cp p}\neq {\cp 0}} \Bigg[ \sum_{i=1}^{3} 
    \Big( E_{i\mathbf{p}} - \epsilon_{i\mathbf{p}} - n_i g_{ii}
    +
    \sum_{\substack{j=1}}^{3}
    \frac{2 m_{ij} g_{ij}^{2} n_i n_j}{{\cp p}^{2}}\Big) 
    - \Omega \frac{n_{2} + n_{3}}{\sqrt{n_{2}n_{3}}}\Bigg].
\end{aligned}
\end{equation}
The total energy density is then
\begin{equation}
    \mathcal{E}(S, n_{1}, n_{23}) = \mathcal{E}_{\text{MF}} + \mathcal{E}_{\text{LHY}}.
\end{equation}
\end{widetext}

Minimizing $\mathcal{E}$ with respect to the spin polarization $S$, we can express $\mathcal{E}$ solely as a function of the densities $n_{1}$ and $n_{23}$. This then determines the chemical potentials \(\mu_{1} = \partial \mathcal{E}/\partial n_{1}\) and \(\mu_{23} = \partial \mathcal{E}/\partial n_{23}\), and a self-bound droplet in vacuum requires 
\begin{equation}
    \mathcal{P}(n_{1}, n_{23}) \equiv \mu_{1} n_{1} + \mu_{23} n_{23} - \mathcal{E}(n_{1}, n_{23}) =0. \label{zero_P}
\end{equation}

Apparently, the single condition in Eq.~(\ref{zero_P}) cannot uniquely determine $n_{1}$ and $n_{23}$. In general, one additionally requires an energy minimum under zero-pressure conditions. Combining this requirement with Eq.~(\ref{zero_P}) yields a unique solution, $n_{1},n_{23}$, for the droplet. Furthermore, to ensure that such a self-bound droplet is stable, one must have
\begin{equation}
  \mu_{1} \le \mu_{1}^{(0)},\ \ \ \mu_{23} \le \mu_{23}^{(0)}, \label{mu} 
\end{equation}
with $\mu_{1}^{(0)} = 0$ and $\mu_{23}^{(0)}=-\sqrt{\Omega^2+\delta^2}$ being the chemical potentials in vacuum (i.e., for a single-particle system). The inequality $\mu_{\alpha} \le \mu_{\alpha}^{(0)}$ ensures that an additional $\alpha$ atom prefers to stay with the droplet rather than in the vacuum, effectively preventing atom loss from the droplet to the surrounding environment. In the limiting case where $\mu_{\alpha} = \mu_{\alpha}^{(0)}$, the droplet and the vacuum reach equilibrium for the $\alpha$ component. This is also a stable solution, corresponding to liquid-gas coexistence as studied previously~\cite{Cui_1, Yu}.

\begin{widetext}

\begin{figure}[h]
\includegraphics[width=\textwidth]{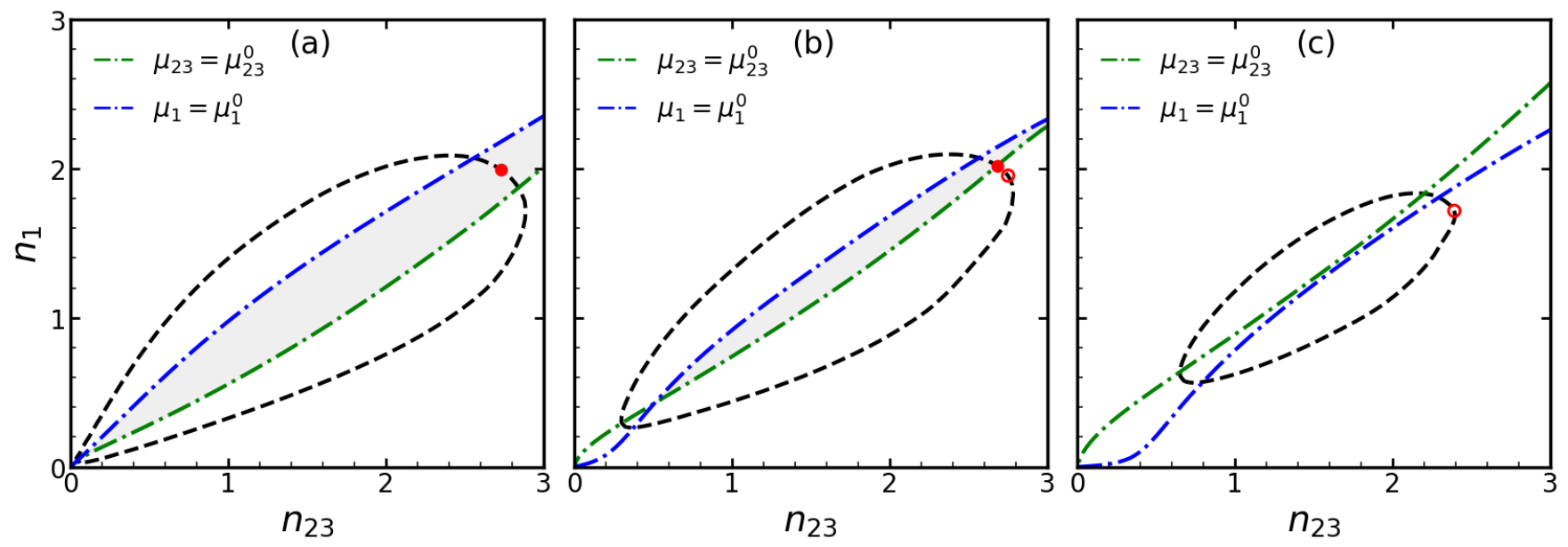}
\caption{Quantum droplet solutions in the $(n_{1},n_{23})$ plane at fixed $\delta=0$ and varying $\Omega/(\hbar \nu_0)=0.2$ (a), $0.8$ (b), and $1.4$ (c), with $\nu_0=20$ kHz. The black dashed curve shows the zero-pressure contour, while the blue and green dash-dotted curves denote the locations where the chemical potentials meet the vacuum values: $\mu_{\alpha}=\mu_{\alpha}^{(0)}$. The shaded region marks the regime where Eq.~(\ref{mu}) is satisfied. The red solid dot indicates the stable droplet solution on the zero-pressure contour. The red hollow dot marks the global minimum on the zero-pressure contour when it lies outside the stable region. In these plots, stable droplets exist in panels (a) and (b) but not in panel (c). The density unit is $\hbar \nu_0/g_{33}$, and the other parameters are the same as those specified in the main text for the Na-Rb system.}
\label{fig_droplet}
\end{figure}

\end{widetext}

In Fig.~\ref{fig_droplet}, we show typical droplet solutions in the $(n_{1},n_{23})$ plane at $\delta=0$ and varying $\Omega$. For small $\Omega$ in Fig.~\ref{fig_droplet}(a), the red solid dot represents the energy minimum on the zero-pressure contour, which lies within the shaded region satisfying the chemical potential condition~(\ref{mu}). It therefore represents a stable droplet solution. As $\Omega$ increases, the shaded area becomes narrower, and the global minimum of the zero-pressure contour can move outside the shaded region [see the red hollow dot in Fig.~\ref{fig_droplet}(b)]. Nevertheless, the zero-pressure contour can still cross the shaded area, and one can identify the stable solution as the lowest-energy point on the contour within the area, as shown by the solid red dot in Fig.~\ref{fig_droplet}(b). Upon further increasing $\Omega$, the shaded area disappears and a stable droplet solution can no longer be supported [see Fig.~\ref{fig_droplet}(c)]. We note that the stable droplet solutions in Figs.~\ref{fig_droplet}(a) and \ref{fig_droplet}(b) all lie in the Collapse-II region in the mean-field phase diagram (Fig.~\ref{fig_diagram}), instead of the Collapse-I region where the LHY energy with low densities is insufficient to stabilize the system from collapse.

\begin{figure}[t]
\includegraphics [width=8cm]{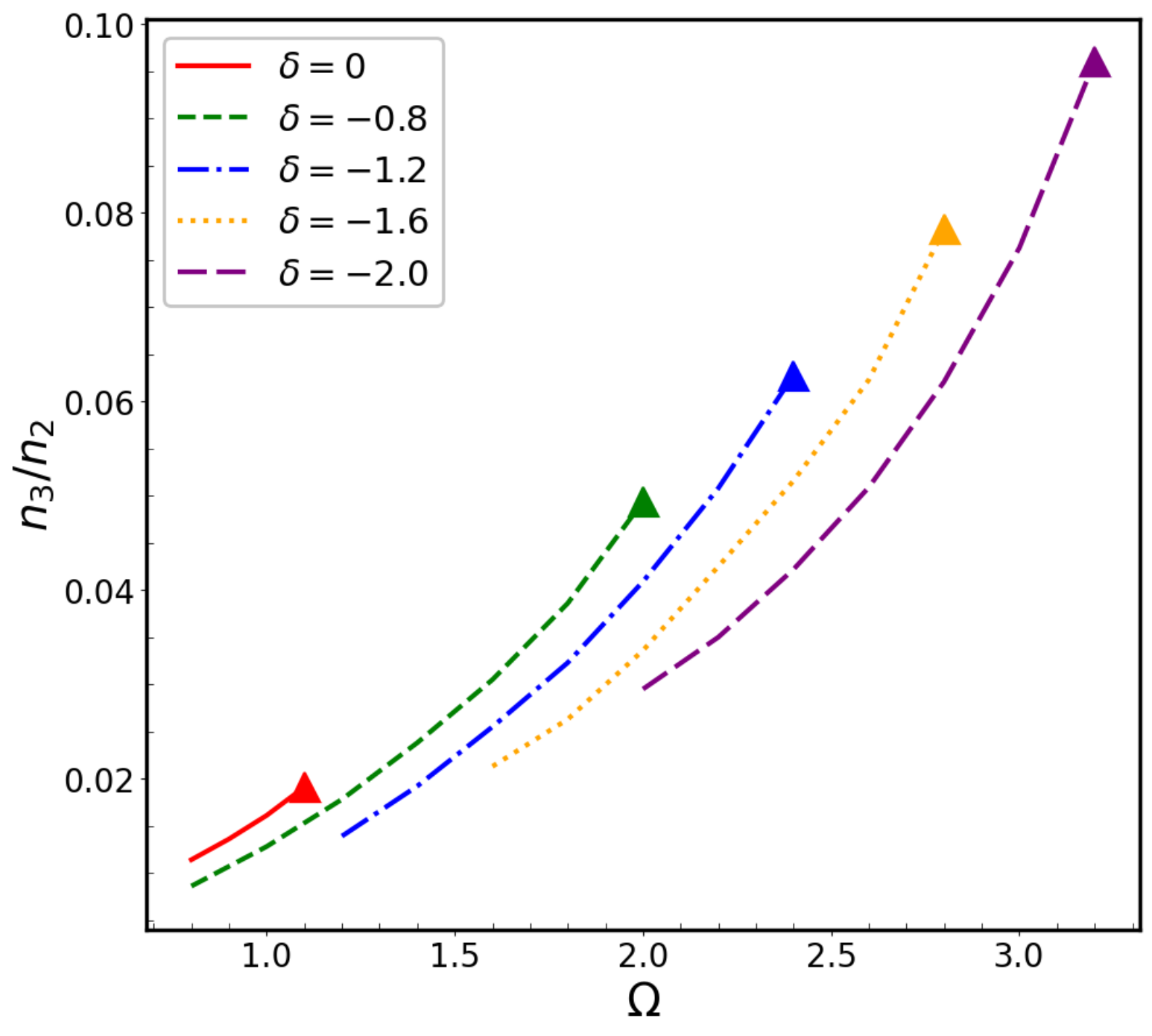}
\caption{Density ratio $n_3/n_2$ as a function of $\Omega$ at different detunings $\delta=0, -0.8, -1.2, -1.6$, and $-2.0$. The triangles mark the upper bounds of $\Omega$ and $n_3/n_2$ for a stable three-component droplet. The units of $\Omega$ and $\delta$ are $\hbar\nu_0$'s, with $\nu_0=20$ kHz. Other parameters are the same as those specified in the main text for the Na-Rb system.}
\label{fig_delta}
\end{figure}

As indicated in Fig.~\ref{fig_droplet}, there is an upper bound on $\Omega$ for achieving stable three-component droplets. Physically, this can be understood as follows. As $\Omega$ increases, components 2 and 3 become more strongly coupled, and therefore more 3 atoms join the droplet, as shown by the increase in $n_3/n_2$ in Fig.~\ref{fig_delta}. Since interactions between component 3 and the other components are all repulsive, the increase in $n_3$ introduces more repulsive forces into the system and eventually destabilizes the self-bound droplet. Consequently, along with the upper bound on $\Omega$, there is an upper bound on $n_3/n_2$ for a stable droplet. As shown in Fig.~\ref{fig_delta}, at $\delta=0$, the upper bound of $n_3/n_2$ is about $2\%$, meaning that the three-component droplet is still dominated by the (1,2) binary droplet, with only a tiny fraction of component 3.

Interestingly, the fraction of component 3 can be efficiently enhanced by switching on a finite detuning $\delta$ between it and component 2. As shown in Fig.~\ref{fig_delta}, when $\delta$ is tuned negative and its magnitude increased, a stable three-component droplet is allowed over a broader range of $\Omega$, enabling a higher fraction of 3 (or $n_3/n_2$) at the upper bound of $\Omega$. For instance, at $\delta/\hbar = -40$ kHz, a stable droplet solution can extend to $\Omega = 64$ kHz, at which $n_3/n_2$ can reach about $10\%$. Such $\delta$-stabilized droplet formation can be understood as follows. When a finite $\delta$ is turned on, the shift in $\mu_{23}$ is dominated by $-\delta S$, as seen from the mean-field energy in Eq. (\ref{E_mf_0}). Given $S<0$ ($n_3<n_2$), this shift is negative for $\delta<0$, meaning that $\mu_{23}$ decreases linearly as $|\delta|$ increases. On the other hand, the vacuum value $\mu_{23}^{(0)}$ decreases quadratically as $|\delta|$ increases. Consequently, at small negative $\delta$, the condition $\mu_{23}<\mu_{23}^{(0)}$ is more easily satisfied compared to the case $\delta=0$. In other words, the shaded area in Fig.~\ref{fig_droplet} (satisfying (\ref{mu})) broadens as $\delta$ becomes more negative, thereby favoring the droplet formation.

So far, we have discussed the properties of three-component quantum droplet in the thermodynamic limit ($N_{i}, V\rightarrow\infty$ with fixed $n_i\equiv N_i/V$). However, in realistic ultracold systems, atom numbers are finite and one must consider finite-size effects. To describe a finite-size droplet, we employ the extended Gross-Pitaevskii (GP) equations under the local density approximation:
\begin{equation}
\begin{aligned}
    i \hbar \frac{\partial}{\partial t} \psi_{1}
    &= \left(
    -\frac{\hbar^{2}\nabla^{2}}{2 m_{1}}
    + \sum_{i} g_{1i} n_{i}
    + \frac{\partial \mathcal{E}_{\rm{LHY}}}{\partial n_{1}}
    \right) \psi_{1},\\
    i \hbar \frac{\partial}{\partial t} \psi_{2}
    &= \left(
    -\frac{\hbar^{2}\nabla^{2}}{2 m_{2}}
    + \sum_{i} g_{2i} n_{i} +\delta
    + \frac{\partial \mathcal{E}_{\rm{LHY}}}{\partial n_{2}}
    \right) \psi_{2}
    - \Omega \psi_{3}, \\
    i \hbar \frac{\partial}{\partial t} \psi_{3}
    &= \left(
    -\frac{\hbar^{2}\nabla^{2}}{2 m_{3}}
    + \sum_{i} g_{3i} n_{i} - \delta
    + \frac{\partial \mathcal{E}_{\rm{LHY}}}{\partial n_{3}}
    \right) \psi_{3}
    - \Omega \psi_{2}.
\end{aligned}
\label{GP}
\end{equation}
with \(n_{i}(\textbf{r}) = |\psi_{i}(\textbf{r})|^{2}\). Based on the coupled GP equations, we have obtained the ground state by performing imaginary-time evolutions. In view of the underlying symmetry, we consider only a spherically symmetric droplet. This configuration allows the largest density overlap between different components and is therefore believed to be energetically favorable over other asymmetric configurations for a miscible three-component state.

\begin{figure}[h] 
\includegraphics [width=0.95\linewidth]{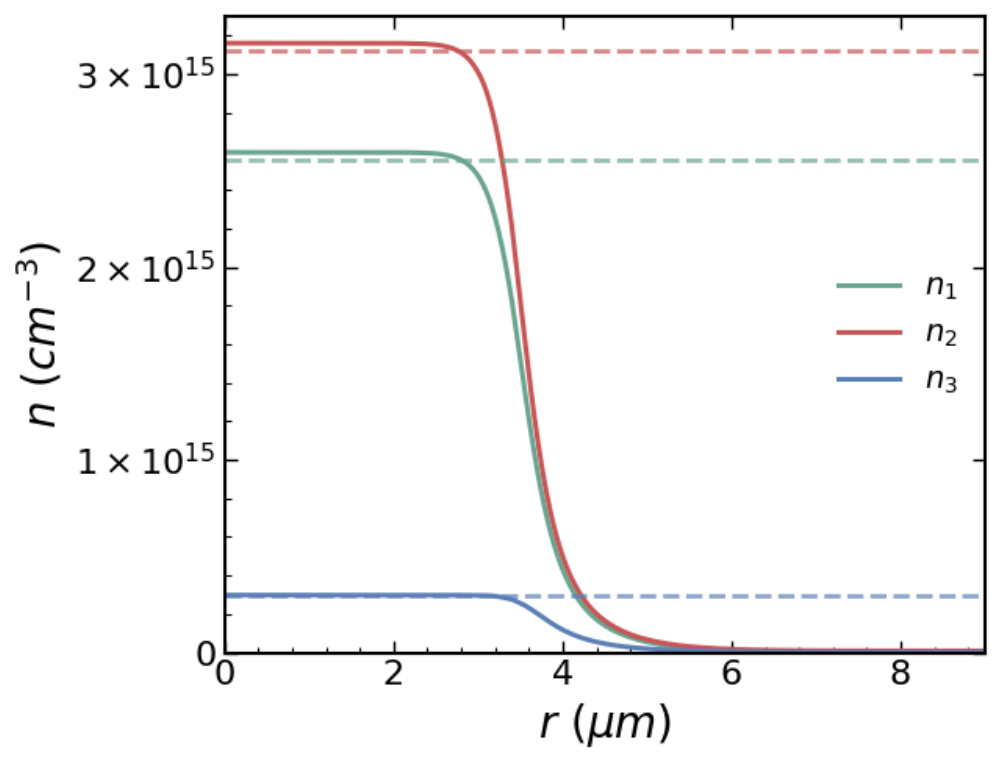}
\caption{Density profiles of a three-component quantum droplet at $\delta=-2$ and $\Omega=3.2$. The units of $\Omega$ and $\delta$ are $\hbar\nu_0$'s, with $\nu_0=20$ kHz. The solid lines are from numerical simulations based on Eq.~(\ref{GP}), with particle numbers $(N_{1},N_{23})/10^5 = (9.1, 6.5)$. The horizontal dashed lines show equilibrium densities in the thermodynamic limit.
} 
\label{fig_GP}
\end{figure}

In Fig.~\ref{fig_GP}, we show typical density profiles of a three-component droplet from GP simulations, corresponding to the highest triangle point in Fig.~\ref{fig_delta}. All three components exhibit flat-top densities in the bulk, which match very well with the equilibrium densities in the thermodynamic limit (see dashed horizontal lines in Fig.~\ref{fig_GP}). We have also tested cases with larger $\Omega$ where the thermodynamic analysis does not predict a stable droplet, as in the example shown in Fig.~\ref{fig_droplet}(c). In those cases, the GP simulations do not yield a stable self-bound solution, as expected. These results verify our theoretical analyses of the stabilization conditions for a three-component quantum droplet.

\section{Summary and Outlook} \label{summary} 

In summary, we have shown that Rabi coupling can facilitate the formation of a three-component quantum droplet even when only one interspecies attraction exists among the three components. Specifically, the Rabi field couples one component of a binary droplet to a third component, thereby gluing all three components together as a self-bound object. Increasing the Rabi coupling leads to a higher fraction of the third component, but also destabilizes the droplet due to the involvement of more repulsive forces. Furthermore, we show that a finite magnetic detuning between the Rabi-coupled components can enhance the stability of the three-component droplet and conveniently tune the upper fraction of the third component. All these results can be readily detected in current cold-atom experiments on Na-Rb mixtures, as well as in other boson-boson mixtures with similar setups of external fields and interatomic interactions.

In this work, all calculations have been performed in a single magnetic field corresponding to a fixed \(a_{12}\). Since the three-component droplet is ultimately stabilized by
the attractive interaction between components 1 and 2, a natural expectation is that making \(a_{12}\) more negative---by tuning the magnetic field closer to the Feshbach resonance---would strengthen the attractive mean-field contribution and thereby allow the system to accommodate a larger fraction of the repulsive component 3 before the droplet is destabilized. This would effectively widen the stability window in \(\Omega\) and enlarge the accessible range of three-component droplet compositions. We have numerically verified this expectation for a value of \(a_{12}\) that is more negative than the present case. A systematic study of how the droplet phase diagram
evolves with \(a_{12}\) remains an interesting direction for future work.

Beyond the three-component framework, our results outline a general route for stabilizing multicomponent droplets. Namely, one can first prepare a basic binary droplet and then couple it to additional components using suitable single-particle fields. Here, the single-particle field serves as a bridge between the existing droplet and the additional components, thereby allowing the formation of larger self-bound objects with more components. On the other hand, a side effect of such single-particle fields is that they may effectively weaken the binding strength of the basic droplet and eventually destroy the self-bound state, as shown in the present work at large Rabi-coupling strengths. In the future, it will be interesting to explore the collective excitations and dynamical formation of such multicomponent droplets, which are expected to reveal the intrinsic interspecies correlations in a more visible way.

\acknowledgments
This work is supported by the National Natural Science Foundation of China (Grants No. 12525412, No. 92476104, and No. 12134015) and the Innovation Program for Quantum Science and Technology (Grant No. 2024ZD0300600). D.W. is supported by the Hong Kong RGC General Research Fund (Grant No. 14304024) and the Collaborative Research Fund (Grant No. C4050-23G).

\section*{Data Availability}
Data that support the findings of this study are openly available~\cite{Cui_8}.

\end{document}